**Computational Modeling of Human Adaptation in Urban Infrastructure Management under Extreme Conditions: A Case Study of Subway Flood Scenarios**

Jinfeng Lou[a], Zijie Liang[b], Pengkun Liu[c], Yuxin Zhang[d], Cleotilde Gonzalez[e], Pingbo Tang[f, *]

[a] Post Doc Fellow, Department of Civil and Environmental Engineering, Carnegie Mellon University, 5000 Forbes Avenue, Pittsburgh, PA 15213, United States

[b] Ph.D. Candidate, Department of Civil and Environmental Engineering, Rice University, 6100 Main Street, Houston, TX 77005, United States

[c] Post Doc Fellow, Hong Kong Center for Construction Robotics, The Hong Kong University of Science and Technology, Hong Kong, China

[d] Research Assistant Professor, Department of Building Environment and Energy Engineering, The Hong Kong Polytechnic University, Hong Kong , China

[e] Professor, Department of Social and Decision Sciences, Carnegie Mellon University, 5000 Forbes Avenue, Pittsburgh, PA 15213, United States

[f] Associate Professor, Department of Civil and Environmental Engineering, Carnegie Mellon University, 5000 Forbes Avenue, Pittsburgh, PA 15213, United States

[*] Corresponding author, email: ptang@andrew.cmu.edu

**Abstract**

Decision-making in urban infrastructure management during extreme events relies heavily on human operators, yet current computational support systems often fail to account for non-monotonic human adaptation and latent psychological biases like overconfidence and defensive overcorrection. This study addresses this gap by integrating Instance-Based Learning Theory (IBLT) into the domain of civil engineering computing. We establish a computational cognitive architecture that simulates operator decision processes through the mathematical mechanisms of memory retrieval and utility blending. This model functions as a computational baseline, representing boundedly rational adaptation driven by experiential priors, thus allowing for the algorithmic isolation of latent psychological biases from the baseline dynamics of memory-based learning. We demonstrated this framework using a human-in-the-loop microworld experiment simulating subway flood-induced track suspensions, where dispatchers must balance passenger safety against service efficiency.

Analysis revealed a complex, non-linear human adaptation cycle consisting of four phases: acquisition, overconfidence, overcorrection, and recalibration. Specifically, the computational model exposed a significant divergence during the post-accident "overcorrection" phase: while human operators exhibited immediate, defensive risk overestimation, the model maintained a stable trajectory based on accumulated experience. This strategic divergence confirms that operational instability following failure is often attributable to acute psychological bias overriding stable memory-based adaptation, a pattern theoretically expected to recur across analogous high-stakes environments and validatable through multi-modal behavioral and sensor data from professional operators. These findings contribute a replicable computational framework to civil engineering computing and lay the groundwork for deploying cognitive architectures as Cognitive Digital Twins within infrastructure control systems, i.e., real-time tools for detecting irrational deviations and informing operator resilience training.

## 1. Introduction

Human decision-making in high-stakes environments is inherently susceptible to systematic psychological biases that can lead to suboptimal behavior (Kunreuther et al., 2002). For instance, periods of successful operation may lead to complacency or overconfidence, causing operators to underestimate accumulating risks (Årstad & Aven, 2017). Conversely, encountering a negative outcome may trigger the "hot stove effect," where decision-makers stop selecting a risky option entirely after a failure, leading to suboptimal risk aversion and a failure to correct internal mental models (Denrell & March, 2001). Without calibrated risk preferences, individuals may engage in defensive over-correction rather than adaptive learning. Because these biases are often latent, empirical observation of performance metrics alone is insufficient to systematically analyze these complex behavioral patterns.

Computational cognitive modeling provides a foundational methodology to address this challenge by developing systems that accurately emulate observed human cognitive constraints (Busemeyer & Diederich, 2010; Gonzalez, 2023). In this study, we utilize the Instance-Based Learning Theory (IBLT) (Gonzalez et al., 2003) to construct a model that simulates decision-making based on memory retrieval, similarity, and decay. This model functions as a computational baseline, simulating how a boundedly rational operator adapts based solely on the retrieval of accumulated experience, free from acute emotional distortions. Comparing human behavior against this memory-based benchmark allows us to dissect the underlying mechanisms of operator actions and identify specific moments where human strategy diverges from stable adaptation. Crucially, this approach helps distinguish between errors caused by incomplete information and those caused by psychological biases.

The need for such diagnostic computational tools is critical in the management of urban infrastructure under extreme conditions. Urban subway systems are increasingly vulnerable to escalating threats (He et al., 2024), as demonstrated by the catastrophic flooding of the Zhengzhou subway (Zhao et al., 2023) and the inundation of New York City's transit network during Hurricane Ida (Mossel et al., 2024). While civil engineers strive to design robust physical infrastructure, the resilience of these systems during emergencies ultimately depends on the real-time decisions of human dispatchers (Prostejovsky et al., 2019). These dispatchers must balance the conflicting mandates of prioritizing passenger safety by preventing trains from becoming trapped in flooded tunnels while simultaneously maintaining service efficiency for the city (American Public Transportation Association [APTA], 2019).

Making these decisions is inherently challenging because the operational environment is uncertain. Dispatchers in a central control room rarely possess a complete picture of underground conditions due to three systemic limitations inherent in civil infrastructure. First, there are significant data gaps. Many networks, such as the New York City subway, lack modern sensor density. This leaves vast stretches of tunnels unmonitored (Dawood et al., 2020; Gkountis & Zayed, 2015). Second, the infrastructure is defined by system complexity. Interconnected components create non-linear cascading effects (Arrighi et al., 2020; Saadat et al., 2020). A single pump failure or signal malfunction can trigger unpredictable consequences across the network. Third, emergencies create severe communication gaps (Burroughs, 2017; Foltz & Brauer, 2012; Peña-Mora et al., 2010). Train conductors on-site and dispatchers in the control room often possess fragmented pieces of information. This leads to conflicting mental models of the evolving hazard.

Consequently, the dispatcher's role represents a complex dynamic decision making (DDM) task where decisions are interdependent and the system state is often opaque (Gonzalez et al., 2017). Dispatchers cannot rely on precise probability tables. Instead, they must fill information gaps by inferring risks based on accumulating experience. Standard infrastructure management protocols often assume that experience leads to linear performance improvement (Ericsson, 2018). The assumption is that a veteran dispatcher is necessarily more reliable than a novice. Research suggests, however, that learning in such stochastic, high-stakes environments is volatile, and accumulated experience does not guarantee expertise (Brehmer, 1980).

Given this background, this study utilizes a gamified microworld environment integrated with computational cognitive modeling techniques to investigate the dynamics of decision-making during simulated subway flood scenarios. By leveraging these computational methods to analyze complex behavioral patterns, we address two primary research questions:

- **RQ1:** How does experience within the game lead to learning and adaptation? Does experience lead to linear improvement, or does it reveal a non-linear cycle of adaptation and over-correction?
- **RQ2:** Can an Instance-Based Learning (IBL) model accurately capture these dynamics? We examine where the model aligns with human behavior and where it diverges to signal discrepancies between the memory-based benchmark and human behaviors.

This study contributes to computing in civil engineering by introducing a dual-agent IBL computational architecture that algorithmically separates performance errors attributable to information limitations from those driven by psychological bias. This establishes a replicable framework for modeling non-linear operator adaptation in high-stakes environments. For infrastructure management practice, the findings identify the post-failure overcorrection phase as a quantifiable cognitive vulnerability and challenge the assumption that operational experience reliably produces expertise in dynamic emergency settings. Building on these findings, the IBL agent is proposed as the computational core of a Cognitive Digital Twin for real-time bias detection and resilience training design.

The remainder of this paper is organized as follows. Section 2 reviews the literature on cognitive biases, Dynamic Decision Making (DDM), and the application of computational cognitive architectures in engineering. Section 3 details the computational methodology, encompassing the software architecture of the microworld simulation, the algorithmic implementation of the IBLT cognitive model, and the data analytics pipeline used to capture behavioral dynamics. Section 4 presents the experimental and validation results, comparing the model's memory-driven predictions against human performance to quantify strategic divergence. Section 5 discusses the implications for integrating such cognitive models into civil infrastructure computing systems, specifically as "Cognitive Digital Twins" for real-time operator monitoring. Finally, Section 6 concludes with a summary of the contributions.

## 2. Literature Review

### 2.1. Psychological Biases, Experience, and Bounded Rationality

The psychological reality of human learning in complex, dynamic environments dictates that performance improvement is rarely monotonic, often characterized by systematic cognitive biases that lead to suboptimal behavior. This departure from prescriptive models of global rationality is formalized by the concept of *bounded rationality*, which posits that decision-makers resort to simplifying strategies due to limitations in computational capacity and information processing (Simon, 1955, 1956). Consequently, instead of maximizing utility, individuals frequently "satisfice," choosing options that are merely satisfactory (Simon, 1979).

Learning derived from experience is inherently prone to distortion because humans base decisions on memory of past outcomes and experienced frequencies, rather than objective statistical probabilities (Gonzalez et al., 2013). These memory-based processes manifest in specific biases that skew decision-making: 1) *Loss Aversion*, where the psychological intensity of negative outcomes is greater than that of equivalent gains, driving behavior toward conservative, risk-averse strategies (Hochman & Yechiam, 2011); and 2) *The Hot Stove Effect*, an adaptive process where experiencing a negative outcome leads to the restriction of subsequent exploration of that option, thereby fostering pessimistically biased beliefs (Denrell & March, 2001). This reliance on information restriction demonstrates how learning can actively impede the selection of objectively optimal choices (Denrell & March, 2001).

Furthermore, the relationship between experience and performance is complex and non-linear. In fast-paced decisions, value biases are exerted dynamically across distinct phases of decision formation, i.e., anticipation, detection, and discrimination, resulting in a complex interplay of countervailing biases rather than a stable process (Corbett et al., 2023). This dynamic tension relates directly to the speed-accuracy tradeoff, where reduced response caution, captured by decreased boundary separation in accumulation models, results in faster, potentially less accurate choices (Myers et al., 2022). Expertise itself offers no complete immunity to these effects, as demonstrated by the paradoxical nature of expertise identified in error detection research for conceptual models (Boutin et al., 2022). Motivational and developmental factors also interfere: less mature students often view simulations merely as assignments to deliver to pass a course, whereas more mature students approach them as opportunities for experiential learning, significantly influencing the perceived value and engagement (Romme, 2004). Finally, internal psychological states, such as trait anxiety, introduce a pessimistic bias by globally increasing the appraisal of negative event probabilities (Stöber, 1997), and this anxiety can selectively disengage neuronal activity essential for effective decision-making (Moghaddam, 2016). These interconnected psychological mechanisms clarify why accumulated experience often fails to guarantee linearly increasing performance.

### 2.2. Dynamic Decision Making (DDM)

The cognitive biases described above are exacerbated in the context of dynamic decision making (DDM). DDM describes the essential process of selecting a sequence of interdependent actions within an environment that evolves continuously over time (Gonzalez et al., 2003, 2017; Lejarraga et al., 2012). Unlike static, one-shot choices, DDM requires agents to consistently evaluate the current state of the environment, maintain situation awareness, and predict potential changes that may occur as a result of both autonomous system dynamics and the agent's own interventions (Donovan et al., 2015). This process requires continuous learning from experienced outcomes to inform future decisions (Gonzalez et al., 2003).

The inherent complexity of DDM stems from several characteristics. Decisions are sequential and necessitate real-life problem-solving skills (Prezenski et al., 2017). The cognitive process involved often breaks down into multiple stages, including: i) goal identification, ii) information gathering, iii) elaboration and prediction, iv) planning, v) decision making and action, and vi) effect control and self-evaluation (Donovan et al., 2015). Furthermore, these environments involve complexity in structure and dynamics, exemplified by critical tasks such as military decision-making, emergency management, and operator roles in industrial processes. Specific areas of application include cyber defense, supply chain inventory management, and the dynamic allocation of resources (Gonzalez et al., 2003; Nguyen et al., 2022). Subway train suspension represents a typical DDM task, as it requires operators to manage a sequence of interdependent decisions (balancing safety and service efficiency) within an underground environment that evolves autonomously and remains largely opaque during extreme weather events.

To effectively study DDM, researchers utilize complex computer simulations, often termed *microworlds* (Gonzalez et al., 2005), which serve as controlled, realistic experimental environments (Donovan et al., 2015; Romme, 2004). The use of simulations allows researchers to expose participants to the full scope of the dynamic decision process (Donovan et al., 2015).

DDM differs fundamentally from traditional normative decision theories (Gonzalez et al., 2017). For instance, prescriptive decision analysis (DA) primarily aims to aid individuals when their expertise is inadequate for a new or difficult situation, often summarized as making a decision "when you don’t know what to do" (Clemen, 2001). In contrast, Naturalistic Decision Making (NDM) typically describes how experts implement familiar, already-made choices in known environments (Klein, 2017). DDM, therefore, focuses on modeling the cognitive processes and adaptive strategies required for learning and control in dynamic, often uncertain, complex systems where outcomes unfold over time (Gonzalez et al., 2003). To rigorously analyze these cognitive processes, a computational framework that simulates human memory constraints is required.

### 2.3. Computational Cognitive Modeling in Simulating and Predicting Human Engineering Behaviors

Computational cognitive modeling offers a rigorous mechanism for integrating human behavior into civil engineering computing. It empirically studies human decision processes by developing computational systems that accurately emulate observed human cognitive constraints and behavior (Lebiere et al., 2013; Malloy & Gonzalez, 2024). This approach is instrumental because it focuses on specifying the decision-making process and the cognitive mechanisms involved into an actionable algorithm, rather than solely aiming for optimal performance (Gonzalez, 2013, 2022; Lebiere et al., 2013). Cognitive models, unlike typical computational algorithms, adhere to human limitations such as bounded rationality, forgetting, and limited attention (Du et al., 2025; Gonzalez, 2013). By formally describing underlying cognitive processes, these models move beyond descriptive statistical analysis to quantify latent cognitive factors, test formal theories, and advance predictive capabilities in complex domains (Cerracchio et al., 2023; Myers et al., 2022).

Instance-Based Learning Theory (IBLT) is a general theory developed to explain how humans make dynamic decisions solely from experience (Gonzalez, 2013; Gonzalez et al., 2003; Lejarraga et al., 2012). IBLT utilizes mechanisms from the Adaptive Control of Thought-Rational (ACT-R) cognitive architecture, particularly its memory system, to simulate learning and decision-making (Du et al., 2025; Gonzalez et al., 2003; Malloy & Gonzalez, 2024). The mechanism by which IBL models process information is threefold:

1. **Instance-Based Knowledge:** Learning is achieved through the storage of discrete experiences, called *instances*. Each instance is typically recorded as a triplet capturing the situation, the decision taken, and the utility or outcome observed (Gonzalez et al., 2003; Lejarraga et al., 2012; Nguyen et al., 2022).
2. **Recognition-Based Retrieval:** When a decision context is presented, the model uses an activation function derived from ACT-R to retrieve relevant past instances from memory. This retrieval process prioritizes instances that are similar to the current situation and reflects human memory characteristics such as the effects of recency and frequency (Gonzalez et al., 2003; Lejarraga et al., 2012; Nguyen et al., 2022).
3. **Utility Blending:** The expected utility of a potential action is calculated via a *blending* mechanism, which sums the outcomes of all retrieved instances, weighted by their cognitive probability of retrieval (Gonzalez et al., 2003; Nguyen et al., 2022). This value reflects the individual's limited experience and memory-based biases rather than objective probabilities (Gonzalez, 2013; Gonzalez et al., 2003).

IBL models exhibit high generality, successfully predicting human behavior and learning curves across a wide range of complexity, from simple repeated binary choice tasks to complex control tasks, cyber defense applications, and phishing detection (Du et al., 2025; Lejarraga et al., 2012; Nguyen et al., 2023; Uttrani et al., 2022). Furthermore, IBL models effectively capture key human cognitive biases, such as loss aversion and exploratory behavior in sequential decision tasks (Lebiere et al., 2013; Malloy & Gonzalez, 2024; Nguyen et al., 2024). The use of IBLT as a cognitive model provides a grounded explanation for dynamic performance and adaptation (Gonzalez, 2013; Lejarraga et al., 2012).

### 2.4. Summary and Research Gap

The review of existing literature establishes three critical premises. First, human learning in safety-critical environments is non-monotonic and prone to systematic biases like the hot stove effect. Second, infrastructure management is a typical DDM task where the state is opaque and interdependent. Third, while computational cognitive models like IBLT have successfully captured human performance in general contexts, they have rarely been deployed as diagnostic baselines within civil infrastructure computing.

A significant gap remains in the lack of computational frameworks capable of isolating psychological biases from skill acquisition in real-time infrastructure operations. Most post-incident analyses in civil engineering rely on static reports, which fail to capture the longitudinal evolution of an operator's strategy during a crisis. There is currently no established computational method to detect the specific moment an operator's decision logic diverges from rational adaptation due to acute stress or overcorrection. This study addresses this gap by leveraging IBL computational architecture. By integrating this cognitive model into a subway flood microworld, we create a "computational control" that allows us to algorithmically separate errors caused by information gaps (which the model captures) from errors caused by psychological bias (which the model isolates), thereby providing a quantitative basis for enhancing operator resilience.

## 3. Methodology

This study utilizes a comprehensive computational framework to analyze operator adaptation, as illustrated in Fig. 1. The methodology begins with the Subway Train Suspension Simulation (Section 3.1), which models the incomplete sensor coverage and non-linear system complexity inherent in flood scenarios. The framework then bifurcates into two parallel data generation streams: the human experiment (Section 3.2), which captures the empirical "learning loops" of interpretation, prediction, and decision-making; and cognitive modeling (Section 3.3), which employs a Dual-Agent IBL Architecture to generate a parallel memory-based computational baseline. Finally, these streams converge in the metric analytics phase (Section 3.4), where performance is assessed using three quantitative measures: Prediction Error (to study skill and bias), Optimal Decision Rate (to evaluate rationality), and Decision Match Rate (to quantify the strategic alignment between the human operator and the computational agent).

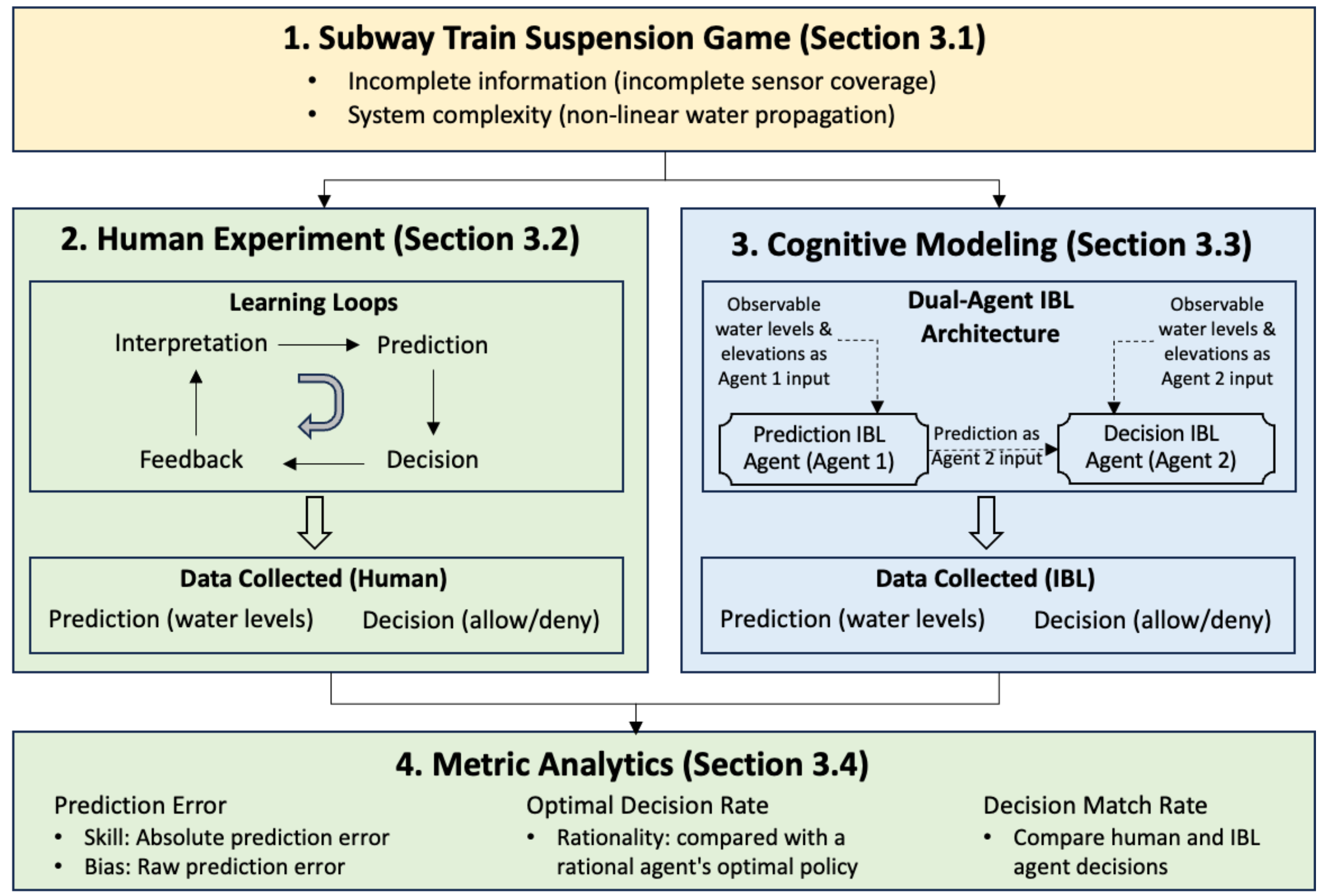

Fig. 1. Overview of the research methodology.

### 3.1. Train Suspension Microworld Game

Studying decision-making during actual subway flood incidents presents significant methodological challenges requiring new computational techniques for investigating human engineering decision and field execution behaviors. Real-world crises are rare and dangerous. It is unfeasible to manipulate these events for experimental purposes. Furthermore, post-incident analyses, such as accident reports, typically reveal only final outcomes. They cannot capture the longitudinal evolution of operator strategies. To plot a learning curve and observe how mental models shift over time, we require a controlled environment augmented by computational engineering processes and cognitive models where the same operator faces repeated, comparable scenarios.

To address these limitations, we adopted a microworld simulation approach. This method isolates specific decision-making variables within a controlled computational environment, allowing for the systematic collection of operator performance data that is typically unavailable in post-hoc accident reports. By standardizing the environmental parameters, we can identify fundamental patterns of human adaptation that generalize to complex field settings.

We developed a web-based simulation titled the "Train Suspension Microworld Game" (Fig. 2) (Lou, 2025). The simulation captures two core challenges: *incomplete information* (incomplete sensor coverage) and *system complexity* (non-linear water propagation). The interface presents participants with a simplified subway network consisting of three nodes: Station A, a central Track Node, and Station B. In each round, the system displays the current water levels at Station A and Station B. Crucially, the water level at the central Track Node is hidden. This simulates the unmonitored areas common in aging infrastructure where sensors are absent.

The environment is governed by a stochastic hydraulic model designed to be learnable but not perfectly predictable. In each round, the system covertly selects a failure point where water influx is determined by a log-normal distribution. Water then propagates to adjacent nodes based on a non-linear function of current water levels and elevation differences. This computationally generated value serves as the objective ground truth for the hidden track node, distinct from historical flood data. Participants assume the role of a dispatcher and perform a dual task in each round:

1. *Prediction:* Based on the visible water levels at the stations and the static elevation data, participants must estimate the water level at the unmonitored Track Node using a slider.

2. *Decision:* Immediately following their prediction, participants must make a binary choice to "Allow Passage" or "Deny Passage."

This design forces participants to construct a mental model of the hydraulic dynamics solely through observation and feedback, mirroring the cognitive load of a real-world operator.

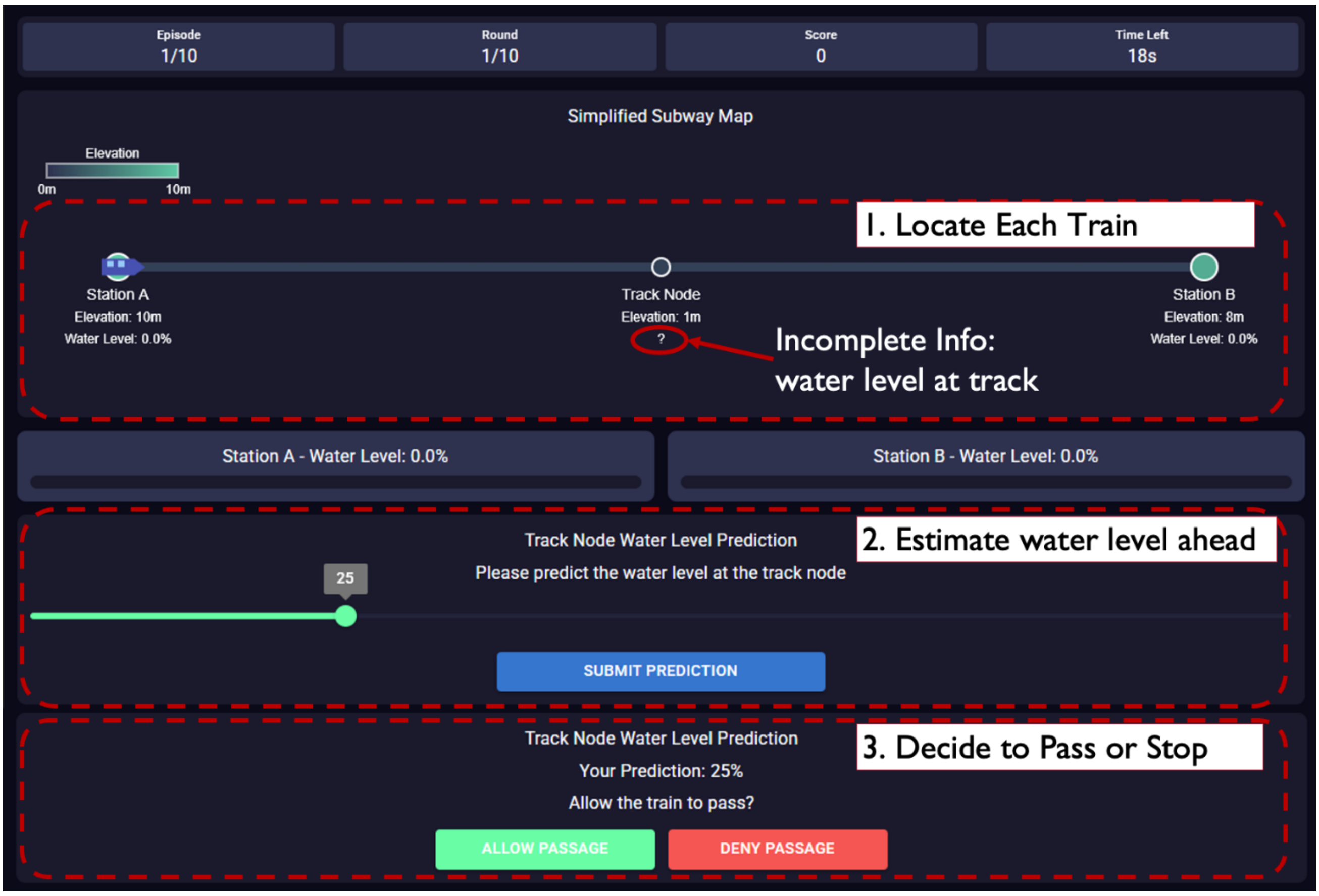

Fig. 2. User interface of the subway flood microworld game

### 3.2. Experimental Design and Procedure

A total of 51 participants were recruited through university mailing lists and the Amazon Mechanical Turk crowdsourcing platform to ensure a sufficient sample size. All participants were required to be at least 18 years of age. The experimental protocol was approved by the Carnegie Mellon University Institutional Review Board (IRB STUDY2025_00000139). To ensure high engagement and effort, a performance-based incentive structure was implemented: the top 50% of participants received a higher compensation (USD 15), while the bottom 50% received a base compensation (USD 5).

The operator's interface displays static information (elevations of all three nodes) and dynamic information (current water levels at Station A and B). To simulate the time pressure of an emergency control room, participants were given a strict 20-second timer to complete both the prediction and decision tasks for each round. The primary objective was to maximize the cumulative *score*, which functions as the "*utility*" in this DDM task. In other words, the game "*score*" serves as the proxy for "*utility*." To simulate the real-world tension between safety and efficiency, the scoring system mapped real-world consequences to discrete utility values and imposed a specific trade-off: a "Trapped Train" (safety failure) resulted in a score loss five times greater than a "Denied Passage" (efficiency loss). This ratio operationalizes a categorical asymmetry that is well-established in infrastructure safety regulation: safety-critical failures and operational inefficiencies are governed by fundamentally different regulatory frameworks, with safety violations incurring consequences that are qualitatively distinct from, and substantially greater than, service disruptions (Reason, 2016; APTA, 2019). The 5:1 ratio reflects this regulatory hierarchy. We acknowledge that the absolute financial and societal costs of major transit incidents (encompassing emergency response, structural recovery, regulatory liability, and reputational damage) may exceed the cost of a preventive service delay by considerably larger factors in specific real-world contexts (Pregnolato et al., 2017). The precise cost ratio in practice varies substantially by jurisdiction, incident severity, and system characteristics, and is therefore difficult to pin down as a single empirical value. The chosen ratio thus represents a conservative operationalization of the safety–efficiency asymmetry, and exploring whether and how the behavioral dynamics observed here shift under different ratio settings is an important direction for future research (see Section 5.4).

While exact real-world costs for safety incidents are difficult to quantify, these relative scores (representing a *50-point penalty* for accidents versus a *10-point penalty* for delays) force the operator to weigh the high magnitude of potential safety losses against the certainty of efficiency losses.

The experiment required approximately 30 minutes and followed a structured sequence (Fig. 3):

1. **Informed Consent:** Participants reviewed a digital consent form detailing the study's purpose, risks, and data confidentiality measures.
2. **Tutorial:** A comprehensive video and text tutorial explained the background information and the scoring criteria.
3. **Gameplay (The Learning Loop):** Participants engaged in 10 distinct episodes.
    - Each episode consisted of a maximum of 10 rounds.

- Crucially, an episode would end prematurely if the participant chose "Allow Passage" when the track node was flooded, resulting in a trapped train. This immediate simulation termination was designed to simulate the catastrophic nature of failure in critical infrastructure.
- Round-level feedback: Immediately following a decision, the system revealed the *actual* water level at the hidden Track Node and the outcome (Success, Trap, or Denied). This immediate feedback loop allowed participants to calibrate their internal predictive models against ground truth.

Informed Consent, Tutorial
purpose, risks, and data confidentiality measures
Gameplay
10 episodes, each with a maximum of 10 rounds
Episode 1
New scenario with new elevations regenerated and water levels starting from 0
Episode 10
Round 1
Water level rises with rounds
Round 2
Water level rises with rounds
Round 10
Early end if trapped before Round 10
Interpretation
water levels & elevations
Prediction
track-node water levels
Learning loops
Feedback
Actual water level, decision results
Decision
Allow or deny

Fig. 3. Longitudinal experimental procedure

### 3.3. The Computational Model: A Dual Agent Architecture

To create a memory-based computational baseline for human performance, we implemented IBLT-based agents or IBL agents. We structured the model using a dual agent architecture to mirror the two-stage cognitive process required by the task:

1. **The Prediction IBL Agent (Agent 1):** This agent simulates the operator's assessment of the environment. In each trial, it observes the current situation (water levels at stations, elevations) and retrieves similar past instances from memory to generate a forecasted water level for the hidden track node. The agent's goal is to minimize the discrepancy between its prediction and the ground truth.
2. **The Decision IBL Agent (Agent 2):** This agent simulates the operator's action. Crucially, its input is not the objective ground truth, but the *output of the Prediction Agent* (Fig. 4). This replicates the real-world condition where dispatchers act based on their prediction of the water level at the unmonitored track node. The Decision Agent retrieves instances of past decisions (Allow vs. Deny) associated with similar predicted water levels to compute a blended value for each option. This blended value functions as a memory-based score estimate, calculated by aggregating the stored scores of retrieved outcomes (e.g., averaging the high penalties of past 'Trapped' instances with the low penalties of 'Deny' instances). The agent then selects the option with the highest blended value. This selection process simulates a risk-based cost-benefit analysis: the agent weighs the potential safety loss of a trapped train (analogous to casualty or property damage costs) against the certain efficiency loss of a denied passage (analogous to economic service disruption costs).

A key architectural feature is the unidirectional information flow between the two agents: Agent 1's prediction feeds into Agent 2's decision, but Agent 2's decision does not feed back into Agent 1's prediction within the same trial. We adopted this unidirectional design for two reasons.

First, the primary purpose of the IBL agent is to serve as a diagnostic baseline representing memory-based bounded rationality without motivational distortion. In high-stress disaster environments, humans may exhibit action-biased cognition, where a preferred decision unconsciously shapes the perception of a hazard. For instance, an operator inclined to allow a train through may unconsciously underestimate the flood risk to justify that choice. A recurrent architecture, in which decision utility feeds back into the prediction within the same trial, would embed this motivational bias directly into the model, eliminating its capacity to detect the bias as a measurable deviation from rational learning. By maintaining a unidirectional structure, any divergence between model predictions and human predictions is attributable to psychological factors, e.g., defensive overcorrection or action-bias, that operate outside the scope of pure memory-based learning.

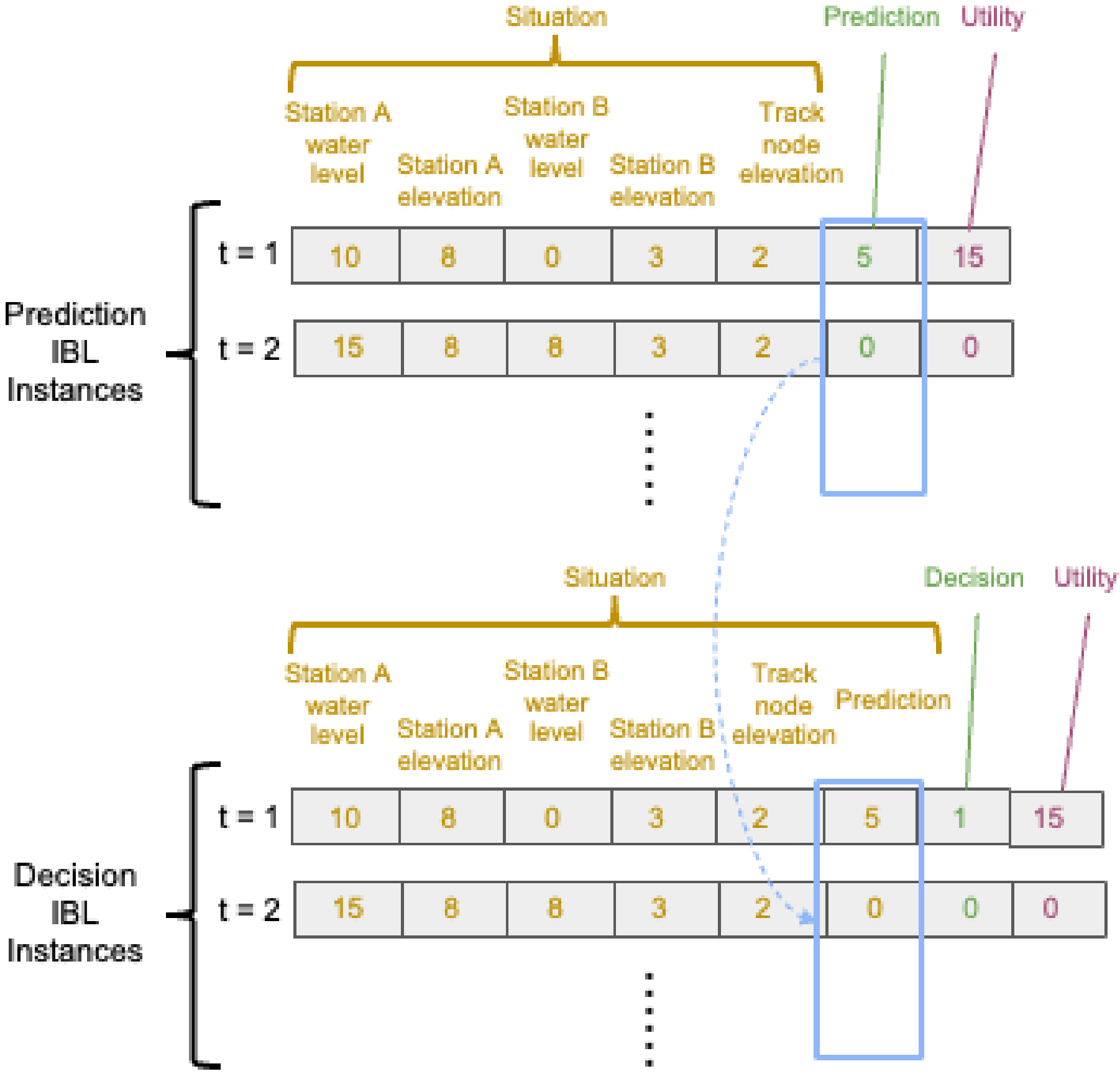

Fig. 4. Data structure of memory instances for the prediction and decision IBL agents

Second, within-trial decision-to-prediction feedback is not required to capture the learning dynamics central to this study. The standard IBL framework already incorporates a cross-trial feedback loop through its instance storage mechanism: after each trial, the observed outcome is recorded as a new memory instance, which subsequently influences future predictions and decisions through retrieval and blending. This cross-trial adaptation is the mechanism through which operators progressively refine their internal models across episodes. The overcorrection phenomenon this study seeks to isolate unfolds at the episode level (across trials, not within them) and is fully captured by this cross-trial feedback without introducing within-trial motivational coupling.

To ensure a valid and robust comparison, we treated the IBL agents as "virtual participants." For every human participant, we ran 30 independent iterations of the corresponding IBL agents through the exact same sequence of trials: the same stochastic water levels, failure points, and feedback chronology as the human counterpart. Each iteration used identical parameters and pre-training initialization but produced a distinct trajectory due to the stochastic noise component ($\sigma$) in the IBL activation function. All reported agent metrics represent the mean across these 30 iterations; the within-participant standard error across iterations was consistently below 2.9%, confirming that the reported patterns are robust to model-internal variance and are not artifacts of a specific random draw.

To bridge the gap between the agent's initially empty memory and the semantic priors humans derive from the tutorial, each IBL agent was initialized through a pre-training protocol in which it was exposed to 5 high-severity simulated episodes (approximately 50 memory instances per agent) drawn from scenarios with rapid water propagation and elevated flood probabilities, simulating the worst-case-scenario exposure received by trained infrastructure operators before assuming operational duties. The Prediction Agent stored instances of the form (observable water levels at Stations A and B, node elevation data, actual water level at the hidden track node), while the Decision Agent stored instances of the form (predicted track water level, chosen action, score outcome). All model parameters were set to established ACT-R defaults confirmed across the IBL literature: decay rate $d = 0.5$, noise $\sigma = 0.25$, and mismatch penalty $\mu = 1.0$ (Gonzalez et al., 2003; Lebiere et al., 2013).

### 3.4. Behavioral and Cognitive Metrics

To analyze the evolution of decision-making and identify points of divergence between human operators and the computational IBL agents, we evaluated three distinct dimensions of operational performance: predictive skill and bias (understanding the hazard), decision rationality (balancing risk and reward), and cognitive alignment (synchronization with the memory-based baseline). Accordingly, we calculated three primary metrics:

***1). Prediction Error (Skill and Bias)***

We utilize two distinct forms of prediction error to disentangle skill acquisition from psychological bias. To benchmark human performance, we calculate these metrics for both the human participants and the Prediction IBL Agent (Agent 1):

*(1) Absolute Prediction Error:* Calculated as $|Prediction_{human,t} - WaterLevel_{actual,t}|$. This metric quantifies the magnitude of water level prediction error. A decrease of such errors over time indicates the acquisition of skill, namely, the operator is learning the physics of the water spread.

*(2) Raw Prediction Error:* Calculated as $(Prediction_{human,t} - WaterLevel_{actual,t})$. This metric reveals directional bias. A positive value indicates overestimation (defensive pessimism), while a negative value indicates underestimation (optimism bias). This is critical for detecting the "over-correction" phenomenon following a failure.

***2). Analytic Optimal Decision Rate (ODR)***

To evaluate the quality of human decisions beyond simple outcomes from the scoring system, we developed a normative benchmark representing a rational agent's optimal policy. This agent maximizes *expected value* (EV), defined here as *the probability-weighted average score of a decision*. By calculating the EV, we can determine the mathematically optimal choice (Allow or Deny) for any given probability of flooding.

Because the game's water generation model is stochastic (due to the random hidden failure point and the log-normal distribution) and non-linear, the objective probability of a trap $P(Trap|Situation_t)$ cannot be calculated analytically. Therefore, we estimated this probability for each trial's initial situation using a Monte Carlo simulation. For the specific situation of each trial (current water levels and elevations), we ran 5,000 simulations of the next round's outcome. This sample size was selected to ensure a robust and statistically stable estimate of $P(Trap|Situation_t)$.

With this objective probability, the EV for each action in trial $t$ was calculated as Eqs. (1) and (2):

$$EV(Allow)_t = P(Trap|Situation_t) * Score(Trap) + (1 - P(Trap|Situation_t)) * Score(Success) \quad (1)$$

$$EV(Deny)_t = Score(Deny) \quad (2)$$

The normatively optimal decision for trial $t$ is the action with the higher EV. We used this benchmark to derive two key performance metrics for each human trial:

- *Optimal decision rate:* A binary measure indicating whether the participant's choice in a given trial matched the choice of the normative model.
- *Optimality gap:* A continuous measure of performance, calculated as the absolute difference between the EV of the optimal action and the EV of the action chosen by the human participant: $|EV(Optimal)_t - EV(Chosen)_t|$. This metric quantifies the magnitude of a suboptimal decision.

***3). Decision Match Rate (Human-IBL Agent Synchronization)***

This metric quantifies the strategic alignment between the human participant and the IBL agent. Since the agent operates as a virtual participant facing the exact same sequence of events as the human, we can compare their choices on a trial-by-trial basis. We calculate the decision match rate for each episode as the percentage of trials where the human and the agent selected the identical decision. The calculation is defined as Eq. (3):

$$DecisionMatchRate_e = \frac{1}{N_e} \sum_{t=1}^{N_e} I(D_{human,t} = D_{agent,t}) \quad (3)$$

In this equation, $N_e$ denotes the total number of rounds in the episode. The variables $D_{human,t}$ and $D_{agent,t}$ represent the binary decisions (Allow or Deny) made by the human and the decision IBL agent at round $t$, respectively. The indicator function $I$ returns a value of 1 if the decisions are identical and 0 otherwise.

This metric serves as an index of synchronization. A high match rate indicates that the human operator is processing information and assessing risk similarly to the instance-based retrieval mechanism of the IBL agent. Conversely, a low match rate signals a divergence in reasoning. This gap allows us to identify specific moments where human decision-making shifts away from the cognitive baseline, potentially due to psychological biases such as defensive over-correction that are not present in the computational agent.

## 4. Results

### 4.1. Human Prediction Skill

This subsection investigates whether participants successfully learned the environmental dynamics of water spreading by analyzing their prediction errors over time. To distinguish between changes in predictive accuracy and shifts in directional bias, we analyze two metrics concurrently: *Absolute Prediction Error* (Fig. 5, blue line), which measures the magnitude of inaccuracy, and *Raw Prediction Error* (Fig. 5, green line), which indicates whether participants were overestimating (positive values) or underestimating (negative values) the water levels.

The initial analysis reveals a significant reduction in Absolute Prediction Error from Episode 1 to Episode 10, indicating that participants successfully learned the fundamental physics of the simulation. However, the learning trajectory is notably non-linear. Rather than a monotonic improvement, the data reveal a pattern characterized by four psychological phases: *Acquisition, Overconfidence, Overcorrection,* and *Recalibration*. To formally test whether performance differed significantly across episodes, we fitted linear mixed-effects models (LMMs) with episode as a categorical fixed factor and participant as a random intercept. All metrics were analyzed at the episode level: for each participant, per-episode means of absolute prediction error, raw prediction error, and optimal decision rate were computed and used as dependent variables. Phase transitions were evaluated via six planned contrasts with Holm-adjusted *p*-values. Effect sizes are reported as partial eta-squared ($\eta_p^2$) for omnibus tests and Cohen's *d* for pairwise contrasts. All statistical results are summarized in Table 1.

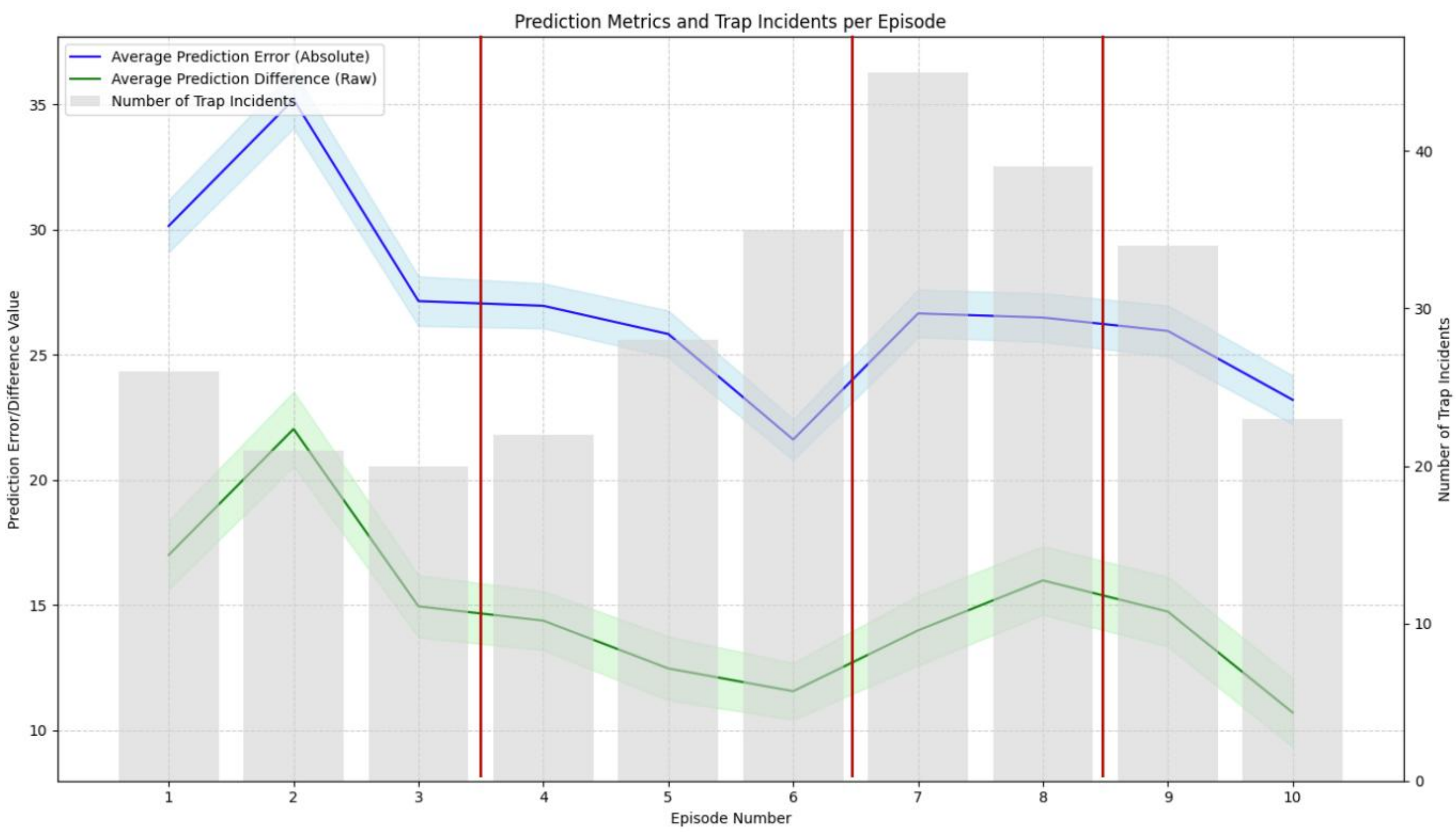

Fig. 5. Human prediction and trap incidents across episodes (Shaded areas: Standard Error (SE))

*Phase 1: Acquisition (Episodes 1–3)*

The first phase is defined by rapid skill acquisition. From Episode 1 (30.2%) to Episode 3 (27.1%), the Absolute Prediction Error shows a net reduction of 3.0 percentage points, albeit with a transient increase to 35.2% in Episode 2 that reflects the initial period of active experimentation as participants probe the system's hydraulic dynamics. Concurrently, the Raw Prediction Error decreases, indicating that participants quickly moved from initial guesswork to a calibrated understanding of the water dynamics. During this period, the number of Trap Incidents (grey bars) remains relatively stable (ranging from 20 to 24), suggesting a cautious exploratory approach where participants focused on understanding system mechanics rather than testing operational limits. Across all ten episodes, the omnibus LMM confirmed a significant effect of episode on absolute prediction error ($p < .001$; Table 1), establishing that performance was not stationary over the course of the experiment.

*Phase 2: Overconfidence (Episodes 4–6)*

By Episode 6, participants achieved their peak predictive performance. The Absolute Prediction Error reached its global minimum (21.6%), and the Raw Prediction Error dropped to its lowest point (11.6%), indicating highly accurate mental models of the environment. However, this period of high accuracy coincided with a concerning behavioral trend: the number of Trap Incidents began to rise steadily, increasing from 22 in Episode 4 to 30 in Episode 6. This divergence suggests a state of *overconfidence*. Having mastered the prediction task, participants likely began to "test the boundaries" of the system, using their precise predictions to take calculated risks that progressively eroded their safety margins.

*Phase 3: Overcorrection (Episodes 7–8)*

The rising risk-taking in Phase 2 culminated in a systemic failure in Episode 7, where Trap Incidents peaked at 44, namely, the highest failure rate in the experiment. This catastrophic feedback triggered an immediate and dramatic reversal in predictive behavior. In Episode 7, the Absolute Prediction Error spiked upward to 26.7%, a regression to near-novice levels. This increase from Episode 6 (21.6%) was statistically significant ($\Delta = +5.1\%$, $p_{adj} = .002$; Table 2).

Crucially, the Raw Prediction Error (green line) mirrored this spike, increasing to 16.0%. This directional shift from Episode 6 (11.6%) was also statistically significant ($\Delta = +4.4\%$, $p_{adj} = .005$; Table 2), confirming that the increase in error reflected a systematic overestimation bias rather than random noise. This increase indicates a shift toward systematic *overestimation*. This suggests that the increase in error was not due to a loss of cognitive skill (forgetting how water spreads), but rather a strategic *over-correction*. Following the "shock" of the trapped trains, participants adopted a defensive bias, perceiving the environment as more dangerous than it objectively was to justify risk-averse decisions.

*Phase 4: Recalibration (Episodes 9–10)*

In the final episodes, participants demonstrated recovery and integration. The Absolute Prediction Error declined again to 23.2% in Episode 10, and the Raw Prediction Error dropped to 10.7%, the lowest bias observed in the experiment. Simultaneously, Trap Incidents decreased significantly to 22. This phase represents *recalibration*: participants shed the excessive defensive bias from Phase 3 and returned to accurate forecasting, but with a reduced failure rate, indicating the formation of a mature, resilient strategy that balances accuracy with appropriate caution.

**4.2. Human Decision Strategy**

We utilize two key metrics: the Number of Trap Incidents (Fig. 6, grey bars), representing catastrophic failure, and the Optimal Decision Rate (ODR) (Fig. 6, purple line), which measures the percentage of trials where the participant's choice aligned with the normative mathematical benchmark.

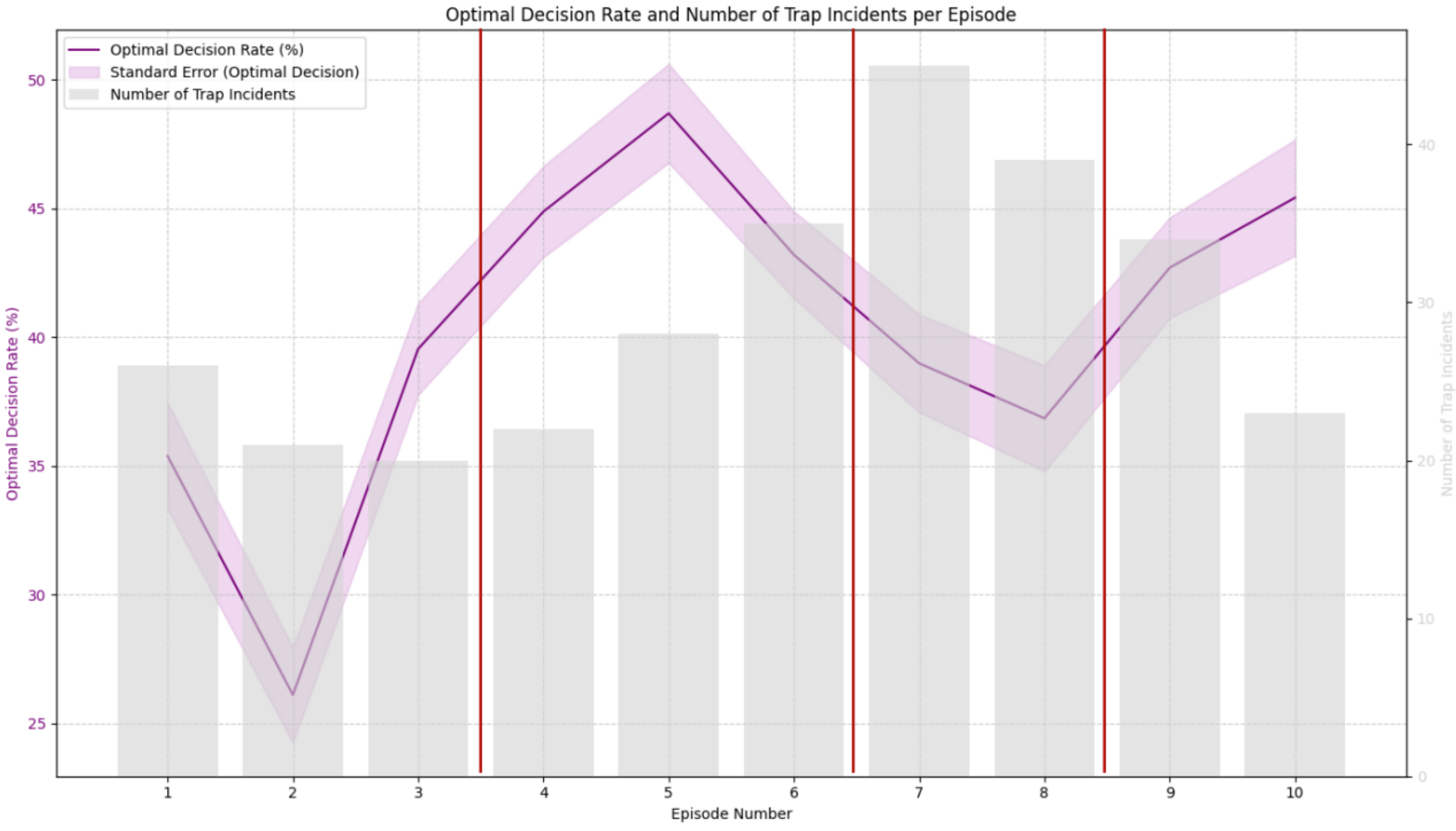

Fig. 6. Human decision optimality and trap incidents across episodes (Shaded areas: SE)

*Phase 1: Acquisition (Episodes 1–3)*

The acquisition phase is characterized by a rapid alignment with rational decision-making principles. After an initial exploratory dip in Episode 2 (where ODR drops to 26%), participants quickly grasped the incentive structure of the task. By Episode 3, the ODR surges, coinciding with the rapid improvement in prediction skill noted in the previous section. During this period, the number of trap incidents remains low, suggesting that participants prioritized safety at the beginning while learning the system's baseline probabilities. However, this risk-averse exploration came at the cost of efficiency; participants frequently denied passage during safe intervals, thereby sacrificing potential system throughput to minimize uncertainty.

*Phase 2: Overconfidence - Boundary Testing (Episodes 4–6)*

In this phase, the relationship between skill and decision quality begins to diverge. The ODR reaches its global peak in Episode 5 (49%), indicating that participants had successfully constructed a mental model capable of identifying high-value opportunities.

However, a critical shift occurs in Episode 6. While prediction accuracy was at its highest (as seen in Section 4.1), the ODR began to decline, and the number of Trap Incidents began to rise noticeably (from 27 in Ep.4 to 33 in Ep.6). This paradox suggests *boundary testing*. Emboldened by their predictive success, participants likely began accepting marginal risks, i.e., choosing "Allow Passage" in situations where the Expected Value was slightly negative or neutral, attempting to "beat the odds." This shift from a strategy of probability maximization to one of risk-seeking laid the groundwork for the subsequent systemic failure.

*Phase 3: Overcorrection - The Shock of Failure (Episodes 7–8)*

The increasing risk tolerance in Phase 2 culminates in Episode 7, where the number of Trap Incidents spikes to 46, the maximum observed in the experiment. This salient negative feedback precipitates a steep decline in decision optimality. The ODR falls steadily, reaching a local valley in Episode 8 (37%). The decline from the peak ODR in Episode 5 (49%) to Episode 8 (37%) was statistically significant ($\Delta = -12.0\%$, $p_{adj} < .001$; Table 2).

Crucially, the nature of suboptimal decisions in this phase differs fundamentally from Phase 2. Whereas the earlier decline was driven by *risk-seeking* (allowing trains in dangerous conditions), the low ODR in Phase 3 is driven by *irrational risk aversion* (over-correction). Traumatized by the spike in traps, participants began choosing "Deny Passage" even in scenarios where the probability of success was high and the expected value was positive. Consequently, operational efficiency plummeted. By systematically rejecting viable transport opportunities, participants incurred significant opportunity costs, effectively trading service reliability for an exaggerated safety margin.

*Phase 4: Recalibration (Episodes 9–10)*

In the final episodes, the decision strategy stabilizes. The Number of Trap Incidents drops sharply (returning to 24 in Episode 10), and the ODR rebounds significantly, rising back to 45%. This recovery mirrors the re-calibration of predictive skill. Having experienced the consequences of both recklessness (Phase 2) and paralysis (Phase 3), participants adopted a balanced, mature strategy. They successfully reintegrated probability assessments into their choices, avoiding the extremes of overconfidence and fear to achieve a sustainable balance between safety and efficiency.

**4.3. Cognitive Modeling**

To explain the behavioral volatility observed in the human data, particularly the regression in performance following systemic failure, we compare the empirical results with those of the IBL agents. As a computational representation of bounded rationality without physiologically biased responses, the IBL agents serve as a cognitive baseline. Deviations between the human and the agents allow us to distinguish between errors caused by limited experience (which both humans and IBL agents suffer from) and errors caused by physiological biases (which only humans experience). All IBL agent trajectories reported below represent the mean across 30 independent iterations per participant. A linear mixed-effects model including an episode × source (human vs. IBL agent) interaction term confirmed that the two prediction-error trajectories diverged significantly across the ten episodes ($p = .012$; Table 3). In Table 4, simple-effects tests revealed that the human–agent gap was non-significant at Episode 6 ($\Delta = 1.9\%$, $p = .155$) but became significant at Episode 7 ($\Delta = 3.9\%$, $p < .012$) and remained significant at Episode 8 ($\Delta = 4.4\%$, $p = .006$), statistically confirming that the divergence is specific to the overcorrection phase.

*Phase 1: Acquisition (Episodes 1–3)*

During the initial episodes, the human and IBL agent trajectories differ due to their distinct initialization states. As shown in Fig. 7, humans begin with lower prediction error than the agent (30.2% vs 34.5% in Episode 1). This indicates that participants utilized semantic knowledge from the tutorial to quickly approximate the water dynamics. In contrast, the prediction IBL agent, which was initialized with instances from high-complexity scenarios to simulate a safety-critical operator, exhibited a more conservative starting strategy. Consequently, the agent required additional rounds to accumulate instances of the specific low-water conditions present in Episode 1. The decision match rate (Fig. 8) rises steadily from 0.39 in Episode 1 to 0.59 in Episode 4. The initial decision match rate is low as a result of the differing initialization states. Guided by the tutorial and the initial low water levels of Episode 1, human participants immediately adopted a strategy of allowing passage to maximize points. However, the IBL agent, pre-trained on high-complexity scenarios, entered Episode 1 with a pessimistic mental model. Relying on memories where "Allowing" frequently led to trapped trains, the agent preferred a "Deny" strategy even when water levels were low. The human prioritized gain (more points) while the agent prioritized safety (avoiding traps). As the agent accumulated new instances of the easier Episode 1 environment, it recalibrated: the decision match rate rose steadily to 0.59 by Episode 4, indicating that the agent "unlearned" its excessive caution and synchronized with the human's strategy.

*Phase 2: Overconfidence (Episodes 4–6)*

In this phase, the human performance and the IBL agent performance are most closely aligned. Human prediction error reaches its global minimum (Fig. 7, blue line), actually surpassing the accuracy of the prediction agent (red dashed line). Simultaneously, the decision match rate reaches its peak at 0.63 (Fig. 8). This high synchronization suggests that the "risk-taking" behavior observed in Section 4.2 was, in this specific window, largely consistent with rational utility maximization. Both the human and the IBL agents identified high-value opportunities to allow trains to pass.

*Phase 3: Overcorrection (Episodes 7–8)*

The critical finding of this study is the directional divergence observed following the spike in trapped trains at Episode 7.

- Prediction trajectories: As illustrated by the green arrows in Fig. 7, the human and agent trajectories move in opposite directions. While the human prediction error spikes upward (reflecting the defensive overcorrection identified in Section 4.1), the prediction IBL agent's error continues its downward trend. The IBL agent, immune to the psychological impact of the failure, treats the negative outcomes simply as new data points, whereas humans react by distorting their predictive model to justify extreme caution.
- Decision decoupling: This divergence in perception leads to a breakdown in strategic alignment. The decision match rate falls significantly to 0.47 in Episode 8 (Fig. 8). This drop quantifies the extent of the psychological compromise: the decision IBL agent continues to recommend decisions based on experienced instances, while humans deviate to an irrational "zero-risk" strategy, denying passage even when the IBL agent deems it safe.

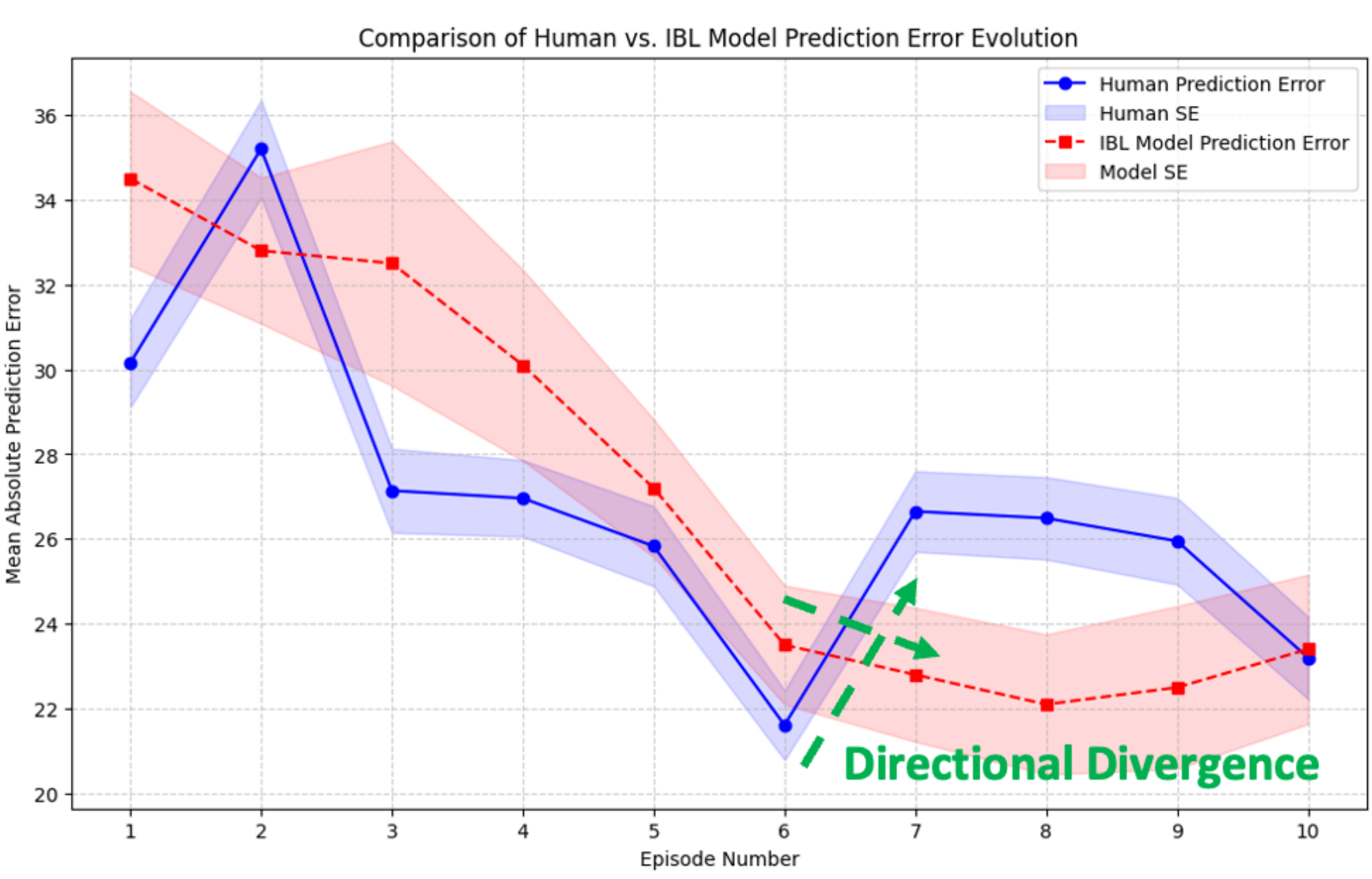

Fig. 7. Comparative evolution of human and IBL agent prediction errors (Shaded areas: SE)

*Phase 4: Recalibration (Episodes 9–10)*

In the final episodes, the human trajectory corrects itself. The human prediction error declines, narrowing the gap with the prediction IBL agent (Fig. 7), and the decision match rate recovers to 0.59 (Fig. 8). This suggests that the psychological interference of the "shock" is temporary. As participants reintegrate the failure event into their broader experience, their decision-making logic re-synchronizes with the memory-based mechanisms of the IBL agent.

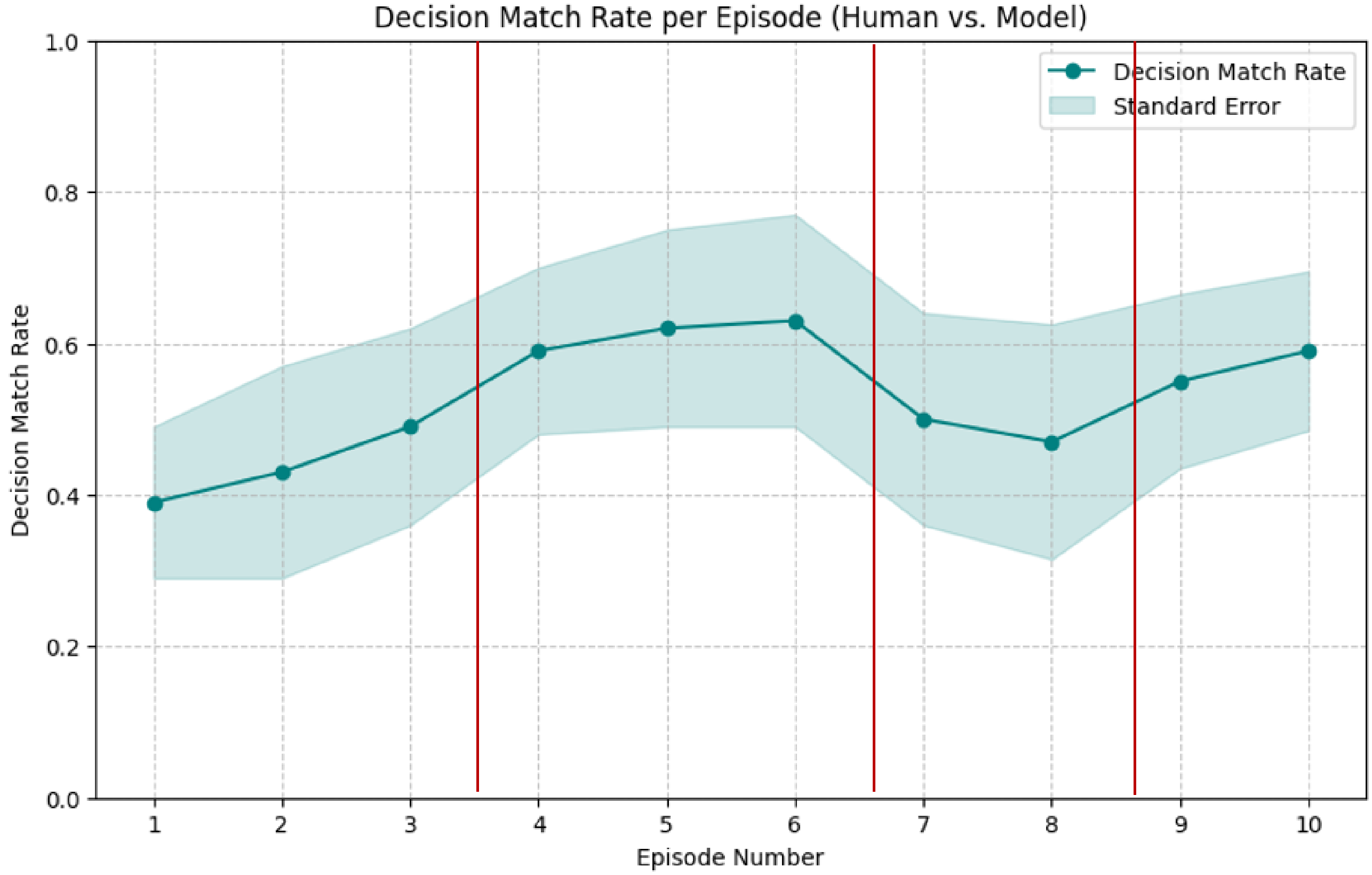

Fig. 8. Decision match rate between human and IBL agent (Shaded areas: SE)

## 5. Discussion

This study investigated the dynamics of decision-making under uncertainty by comparing human behavior against a computational cognitive model created based on IBLT. The findings reveal that experiences in high-stakes environments do not guarantee a linear improvement in performance. Instead, decision-making evolves through a non-linear cycle of adaptation, confidence, failure, and recalibration.

### 5.1. The Cycle of Non-Linear Learning

The behavioral data challenge the assumption that accumulated practice inevitably leads to expertise. While participants successfully learned the fundamental mechanics of the system during the initial episodes, their subsequent performance followed a non-linear trajectory characterized by four distinct phases. After the initial acquisition of skill, participants entered a phase of overconfidence, where high predictive accuracy led to increased risk-taking and a reduction in safety margins. The transition from acquisition to overconfidence mirrors the cognitive shift from "System 2" processing (deliberate, effortful) to "System 1" processing (automatic, heuristic) described in classical learning frameworks. Learners often fail to anticipate the speed of this shift, leading to early overconfidence (Billeter et al., 2011). This behavior culminated in systemic failure, namely, the trapping of trains. Crucially, the immediate human response to this failure was not a rational adjustment of risk parameters, but a defensive over-correction (Metcalfe, 2017). Participants exhibited a strategic regression, characterized by the systematic overestimation of water levels and the rejection of valid, positive-value operational opportunities. This pattern suggests that in dynamic environments, the learning process is cyclical rather than monotonic: operators move from learning to risk-taking, experience a shock from failure, react with excessive caution, and finally recalibrate to a mature strategy. This cycle underscores the fundamental trade-off in infrastructure operations: the 'Acquisition' and 'Overcorrection' phases are characterized by high safety but suboptimal efficiency, whereas the 'Overconfidence' phase maximizes throughput at the expense of catastrophic risk. Understanding this cycle is essential for interpreting operator behavior, as a decline in efficiency metrics after an accident may reflect psychological defense mechanisms rather than a loss of technical competence.

**5.2. The IBL Agent as a Diagnostic Tool**

The comparison between human operators and the IBL agents demonstrates the utility of cognitive modeling as a diagnostic baseline. Because the IBL agent simulates bounded rationality (including memory decay and retrieval limitations, without physiological stress) and was initialized with a prudent, safety-oriented memory set, it represents how a trained, consistent operator would perform if they relied solely on cognitive information processing.

Therefore, the divergence between the human and the IBL agent is a critical data point. During the phases of acquisition and stable performance, human decisions aligned closely with the agent, indicating rational processing. However, following the spike in failures, the decision match rate dropped significantly. This divergence signals that the human operator is no longer relying purely on cognitive retrieval mechanisms. Instead, their decision-making process is likely being distorted by internal psychological factors, such as acute stress or fear of recurring failure. By quantifying this gap, we can distinguish between errors caused by a lack of information (which the IBL agent would also make) and errors caused by psychological interference (which only the human makes).

The diagnostic power of this approach is predicated on the unidirectional architecture of the dual-agent model. Because Agent 2's utility assessments do not influence Agent 1's water-level forecasts within a trial, the model provides a motivationally clean baseline against which human prediction biases become visible. The divergence observed during the overcorrection phase, where human prediction errors spike upward while the model's errors continue to decline, is precisely the signature of action-biased cognition described by Nickerson (1998): operators conditioned by the shock of failure may systematically overestimate flood risk to justify an aversion to allowing trains through. A recurrent model in which decision preferences feed back into predictions could reproduce a portion of this human trajectory, but it would absorb the bias rather than expose it. Future research could develop a complementary recurrent architecture designed specifically to model the magnitude and persistence of motivational interference in operator predictions, rather than to detect it as a deviation from rational learning.

### 5.3. Integration into Civil Infrastructure Computing Systems

The behavioral patterns identified in this microworld study suggest a conceptual framework for potential applications in critical infrastructure management, which would require further validation with professional operators before operational deployment.

*Real-time Decision Monitoring*. We propose deploying the IBL agent as a "Cognitive Digital Twin" within the centralized infrastructure control system. Functioning as a parallel computational process, this module analyzes real-time sensor data streams simultaneously with the human dispatcher. Rather than automating control, the system operates as a diagnostic safeguard: if the human operator's decisions diverge statistically from the model’s rational baseline, specifically if the synchronization rate drops below a calibrated threshold following a critical event, the system generates a bias alert. This prompt serves as a decision aid, signalling potential psychological overcorrection and encouraging the operator to re-evaluate their risk assessment against the memory-based benchmark.

*Targeted Training and Calibration***.** Current training protocols often emphasize error avoidance. However, these findings suggest that operators must be specifically trained to handle the psychological aftermath of failure. Training simulations should intentionally induce catastrophic failure scenarios (similar to the high-failure rate observed in Episode 7) to expose operators to the shock of adverse outcomes. By providing immediate, corrective feedback during these simulated crises, training can help operators recognize their tendency to over-correct and practice maintaining rational risk assessment strategies even immediately after a negative event.

**5.4. Limitations**

Several limitations of this study should be noted. First, the use of a web-based microworld limits the external validity of the findings in two related respects. The financial incentives provided motivation, but cannot replicate the immense psychological pressure and ethical weight of protecting human life in a real-world subway flood; consequently, the defensive over-correction observed in actual emergencies may be even more pronounced than what was observed here. Furthermore, the cognitive patterns identified in this study have not been directly validated against behavioral data from real subway incidents. Such validation remains methodologically challenging given the rarity of well-documented flood events with trial-level dispatcher records, the absence of controlled conditions in post-incident reporting, and the difficulty of isolating individual cognition from organizational response. Future research should seek validation through multi-modal data analysis (integrating trial-level behavioral decision records, real-time system sensor logs, physiological stress indicators, and post-incident reports) to capture the operational complexity that the microworld necessarily simplifies. High-fidelity simulator studies with professional dispatchers represent the most feasible near-term pathway toward this goal.

Second, the participant pool consisted of laypeople rather than professional train dispatchers. While basic cognitive mechanisms regarding risk and memory are universal, experienced professionals may possess specialized heuristics or higher stress tolerance that could moderate the observed effects. Future research should validate these cognitive models using data from professional operators in high-fidelity simulators.

Third, the current experimental design places decision-making authority solely on a single operator. In actual subway control centers, critical decisions such as service suspension typically involve hierarchical approval processes and collaborative information sharing among multiple stakeholders.

This study isolates the individual cognitive process and does not account for the organizational dynamics of distributed decision-making. Additionally, the unidirectional architecture assumes that prediction and decision are cognitively sequential within each trial (an assumption that may not hold under extreme stress), where motivational biases can create bidirectional interference between hazard assessment and action selection (Nickerson, 1998). Developing a recurrent architecture that explicitly models this interference is a direction for future work.

Fourth, this study focused on mean prediction errors to identify phases of adaptation. However, preliminary observations suggest a non-symmetric error distribution characterized by a "long tail" of overestimation, particularly during the overcorrection phase. Future research should specifically investigate these distributional asymmetries to quantify the persistence and magnitude of defensive pessimism. Fifth, this study analyzed the aggregate behavior of the participant pool to identify general phases of adaptation. It did not account for heterogeneity in intrinsic risk tolerance. Individual psychological traits, such as trait anxiety or pre-existing attitudes toward safety, may cause certain operators to diverge from the computational baseline more drastically than others. Future research should integrate psychometric profiling to determine how different operator "types" experience the overcorrection phase, allowing for more personalized calibration of the Cognitive Digital Twin.

Finally, the present study employed a fixed penalty ratio of 5:1 between safety failures and efficiency losses. As noted in Section 3.2, the real-world cost asymmetry in transit operations may differ from this value considerably depending on incident severity, jurisdiction, and system characteristics. Whether the quantitative properties of the observed learning cycle, such as the magnitude of the overcorrection spike or the speed of recalibration, are sensitive to different penalty ratio settings is therefore an open empirical question. A quantitative sensitivity analysis under alternative ratios within this dataset is not methodologically feasible for two reasons. (i) The IBL agents in this framework are paired models calibrated on each participant's behavioral history under the 5:1 incentive structure; re-running them under a different penalty assumption without corresponding human data collected under those conditions would produce a mis-specified model. (ii) The Optimal Decision Rate is defined relative to the incentive structure participants actually faced; applying a normative threshold derived from a different ratio to existing behavioral data would constitute an invalid comparison. Future research employing multi-condition experiments, in which participants are assigned to different penalty ratio conditions, would be required to directly characterize how these behavioral dynamics generalize across the spectrum of real-world cost structures, from moderate regulatory asymmetries to extreme liability environments.

**6. Conclusion**

This study highlights that the resilience of urban subway systems depends not only on physical robustness but also on the cognitive stability of the human dispatchers who manage them. By simulating the uncertainties inherent in aging infrastructure, specifically data gaps and system complexity, we demonstrated that operator performance does not improve linearly with experience. Instead, decision-making in high-stakes environments follows a dynamic cycle characterized by acquisition, overconfidence, overcorrection, and recalibration.

The integration of IBLT provided a critical analytical tool for understanding this cycle. By establishing a memory-based computational benchmark, we were able to isolate the specific moments when human operators diverged from IBL agents. The findings confirm that the drop in operational efficiency following a crisis is often driven not by a lack of competence, but by a psychological overcorrection where operators prioritize excessive caution over system functionality. For the domain of civil infrastructure computing, these findings suggest the value of a shift from purely physical monitoring to integrated socio-technical modeling. If validated with data from professional operators in realistic settings, embedding cognitive architectures into control systems could provide a computational safeguard capable of detecting irrational deviations in real-time. Furthermore, training protocols may benefit from updates to desensitize operators to the shock of negative outcomes, ensuring that they can maintain rational risk assessments even in the aftermath of a crisis. Ultimately, understanding and modeling the non-linear dynamics of human adaptation is essential for designing control systems that remain resilient under the pressures of extreme weather events.

**Data Availability Statement**

Some or all data, models, or code that support the findings of this study are available from the corresponding author upon reasonable request.

**Acknowledgments**

This study was supported by the Carnegie Bosch Institute through the Carnegie Bosch Fellowship.

**Table 1.** Omnibus tests for performance metrics across episodes

| **Metric** | **F** | ***df₁*** | ***df₂*** | ***p*** | $\eta_p^2$ |
|---|---|---|---|---|---|
| Absolute Prediction Error | 8.34 | 9 | 450 | < .001 | .14 |
| Raw Prediction Error | 5.67 | 9 | 450 | < .001 | .10 |
| Optimal Decision Rate | 6.12 | 9 | 450 | < .001 | .11 |
| Decision Match Rate | 7.45 | 9 | 450 | < .001 | .13 |

*Note:* * indicates *p* < .05.

**Table 2.** Planned contrasts for behavioral phase transitions

| Contrast | ΔM | SE | *t* | $p_{adj}$ | Cohen's $d_z$ |
|---|---|---|---|---|---|
| *Absolute Prediction Error:* | | | | | |
| Acquisition (Ep1 → Ep3) | −3.0% | 1.06 | −2.84 | .010 * | 0.38 |
| Peak Skill (Ep3 → Ep6) | −5.5% | 1.71 | −3.21 | .007 * | 0.44 |
| Overcorrection (Ep6 → Ep7) | +5.1% | 1.39 | 3.67 | .002 * | 0.51 |
| Sustained Shift (Ep6 → Ep8) | +4.9% | 1.28 | 3.83 | .002 * | 0.54 |
| Recalibration (Ep8 → Ep10) | −3.3% | 1.13 | −2.92 | .010 * | 0.41 |
| Overall Learning (Ep1 → Ep10) | −6.9% | 1.67 | −4.12 | .001 * | 0.57 |
| *Raw Prediction Error:* | | | | | |
| Overcorrection (Ep6 → Ep7) | +4.4% | 1.37 | 3.22 | .005 * | 0.45 |
| Recalibration (Ep8 → Ep10) | −5.3% | 2.11 | −2.51 | .015 * | 0.35 |
| *Optimal Decision Rate:* | | | | | |
| Peak (Ep3 → Ep5) | +7.2% | 2.49 | 2.89 | .011 * | 0.40 |
| Overcorrection (Ep5 → Ep8) | −12.0% | 2.64 | −4.55 | < .001 * | 0.63 |
| Recalibration (Ep8 → Ep10) | +8.0% | 3.42 | 2.34 | .023 * | 0.33 |

*Note:* * indicates $p < .05$.

**Table 3.** Interaction effects between human operators and IBL agents

| **Metric** | $F_{interaction}$ | $df_1$ | $df_2$ | $p$ | $\eta_p^2$ |
|---|---|---|---|---|---|
| Absolute Prediction Error | 4.23 | 9 | 900 | .012 * | .04 |

*Note:* * indicates $p < .05$.

**Table 4.** Human-agent simple effects at key episodes

| Episode | Δ (human − agent) | $t$ | $p$ |
|---|---|---|---|
| Episode 6 | −1.9% | 1.44 | .155 |
| Episode 7 | +3.9% | 2.60 | .012 * |
| Episode 8 | +4.4% | 2.84 | .006 * |

*Note:* * indicates $p < .05$.